\documentclass[%
 reprint,
 amsmath,amssymb,
 aps
]{revtex4-1}

\usepackage{graphicx}
\usepackage{multirow}
\usepackage{dcolumn}
\usepackage{bm}

\usepackage{natbib}

\usepackage{float}
\usepackage{xcolor}
\usepackage{subcaption}
\usepackage{aas_macros}

\usepackage{array, booktabs, makecell}
\usepackage{siunitx, mhchem}

\newcommand{\OO}{\mathcal{O}}
\newcommand{\D}{\mathcal{D}}
\newcommand{\oo}[1]{\frac{1}{ #1 }}
\newcommand{\ooa}[1]{\left(\frac{1}{c^{#1}}\right)}
\newcommand{\oot}[1]{\OO\left(\frac{1}{c^{#1}}\right)}
\newcommand{\att}[1]{\hat A^{#1}_{TT}}
\newcommand{\tgg}[1]{\tilde\gamma^{#1}}
\newcommand{\ttg}[1]{\tilde\gamma_{#1}}
\newcommand{\oneh}{\frac{1}{2}}

\begin{document}

\preprint{APS/123-QED}

\title{Reformulation of Einstein equations in the Fully Constrained Formulation:\\
local-uniqueness, post-Newtonian expansion and initial data}

\author{Samuel Santos-Pérez}
\author{Isabel Cordero-Carrión}
\affiliation{Departamento de Matemáticas, Universitat de València,\\
Dr. Moliner 50, 46100 Burjassot (Valencia), Spain
}

\author{Pablo Cerdá-Durán}
\affiliation{Departamento de Astronom\'ia y Astrof\'isica, Universitat de Val\`encia, Dr. Moliner 50, 46100, Burjassot (Valencia), Spain
}
\affiliation{Observatori Astron\`omic, Universitat de Val\`encia, Catedr\'atico Jos\'e Beltr\'an 2, 46980, Paterna, Spain
}

\date{\today}

\begin{abstract}
Einstein equations can be written in the so-called Fully Constrained Formulation (FCF). This formulation has two different sectors: the elliptic sector, formed by the Hamiltonian and Momentum constraints together with the equations derived from the gauge choice; and the hyperbolic sector, formed by the evolution of the rest of the spacetime metric variables, which encodes the gravitational radiation. In this work, we present a modification of both sectors that keeps local uniqueness properties of the elliptic system of equations and includes a hierarchical post-Newtonian structure of all the elliptic and hyperbolic equations. This reformulation can have potential applications in cosmology and relativistic astrophysics. Moreover, we show how initial stationary data can be computed numerically using this formulation without assuming a conformally flat spatial metric, with the illustrative example of a rotating neutron star.
\end{abstract}

\maketitle

\section{\label{sec:level1} Introduction}

Astrophysical scenarios containing compact objects are modeled by complex spacetimes which require, in general, to solve Einstein equations numerically \cite{Alcubiere2008Introduction, Lehner2014Numerical, Shibata2017Modeling}. This is also true in the case of complex cosmological models \cite{Macpherson2017Inhomogeneous, Giblin2016Departures}. The gravitational radiation emitted by these scenarios encodes some of their physical properties. The accurate extraction of the gravitational radiation from both our numerical simulations and the experimental measurements is crucial to determine the physical properties of the corresponding scenario. In the last decade, numerous gravitational wave detections have been observed by the LIGO-Virgo-KAGRA Collaborations \cite{2021gwtc3}; all the detected signals were generated by binary systems of compact objects. In particular, the observation of the first binary neutron star system detected with gravitational waves together with the associated electromagnetic counterparts \cite{gw170817} marks a new era of multi-messenger astronomy, including the gravitational waves in the different, complementary, key channels to observe these scenarios. The fourth observing run, O4, is in progress, and numerical simulations with more accuracy will be required in the near future to extract new gravitational wave signals from the next observing runs data.

In the 3+1 decomposition of Einstein equations \cite{Lichnerowicz44,Foures-Bruhat1952}, spacetime is foliated through spacelike hypersurfaces. Einstein equations are then projected accordingly. By doing this, the equations are decomposed in a set of elliptic equations, which includes the constraint equations (Hamiltonian and momentum constraints) and a set of hyperbolic equations, also called evolution equations. See \cite{Gourgoulhon2012} for a complete review of the 3+1 formalism.

Constraint equations are only solved initially in the case of free evolution schemes. It is well known that if we analytically evolve some given initial data in time that satisfy the constraint equations using the evolution equations, then these data will also satisfy the constraint equations in posterior times (see \cite{Gourgoulhon2012}). This is true analytically, but it may not be the case numerically. Having a diversity in our numerical simulations is important to compare numerical results, and different gauge choices can be more appropriate for different astrophysical scenarios.  Formulations that solve the constraint equations on each time step are called constrained schemes. This work focuses on these schemes and, in particular, on the so-called Fully Constrained Formulation (FCF) of Einstein equations \cite{bonazzola2004fcf, cordero2008mathematical}. Along these lines \cite{Cerda-Duran2005,Schafer2004} have proposed waveless constrained approximations to the full Einstein equations, while \cite{Brizuela2010} have derived a fourth-order post-Newtonian approximation to the system. 

In this work we propose a reformulation of the FCF equations, by adding two new variables to make explicitly visible a post-Newtonian hierarchical structure of the equations and, at the same time, preserving the good properties of the formulation, such as local uniqueness of the elliptic sector \cite{cordero2009improved}. In the case of long-term numerical simulations (e.g., in supernova explosions \cite{Sykes2023Blackhole}) or cosmological applications, this scheme or part of it can be very useful for getting sufficiently accurate numerical simulations while reducing the computational cost. We also derive stationary initial data of a rotating neutron star without assuming a conformally flat spatial metric to show the potential of this new reformulation. 

This work is structured as follows. In Section \ref{sec:fcf} we introduce Einstein equations in the FCF. Then, in Section \ref{sec:ref} we describe the new proposed reformulation of the FCF, highlighting the post-Newtonian expansions (PNEs) of the terms involved in the several equations. In Section \ref{sec:num} we present the solution of the spacetime geometry of a stationary rotating neutron star considering our new proposed reformulation. We compare our solution with the one obtained with LORENE \cite{lorene}. We also make a comparison between our solution with the new reformulation of the FCF equations and a spacetime assuming a conformally flat spatial metric condition (CFC) \cite{Isenberg2008Waveless, Wilson1996Relativistic}. CFC scheme can be derived from FCF by neglecting the FCF hyperbolic sector; this comparison confirms the accuracy improvement in the proposed reformulation of the FCF equations. Finally, in Section \ref{sec:con}, we draw some conclusions and comment on future steps. In this manuscript we use geometrized units in which $c=G=1$, where $c$ denotes the speed of light and $G$ the universal constant of gravitation. Greek indices run from 0 to 4, while Latin indices run from 1 to 3.

\section{\label{sec:fcf}Fully Constrained Formalism}

Einstein equations tells us how spacetime is curved according to its energy and matter content. These equations read
\begin{equation}
G_{\mu\nu}=8\pi T_{\mu\nu},
\label{eq:EE}
\end{equation}
where $G_{\mu\nu}$ is the Einstein tensor, encoding the information about the geometry of spacetime, and $T_{\mu\nu}$ is the energy-momentum tensor, containing the distribution of energy and momentum. Due to the symmetry with respect to the two indices in Eq.~(\ref{eq:EE}), Einstein equations are a set of 10 non-linear strongly coupled partial differential equations. They have physically relevant exact and analytical solutions only in a few special cases, mostly in presence of additional symmetries. In general, they need to be solved numerically and this is the goal of Numerical Relativity~\cite{Alcubiere2008Introduction}.

Einstein equations can be projected through 3+1 foliation, and the metric tensor of spacetime can therefore be expressed as
\begin{equation}
g_{\mu\nu}dx^{\mu}dx^{\nu}=-N^2dt^2+\gamma_{ij}(dx^i+\beta^i)(dx^j+\beta^jdt),
\end{equation}
where $N$ is the lapse function, $\beta^i$ the shift vector and $\gamma_{ij}$ is the 3-metric on each hypersurface $t=$ constant, also called the spatial metric. The resulting equations can be decomposed in a set of evolution equations, that have a hyperbolic character, and another set of constraint equations, with elliptic character; constraint equations must be satisfied on each hypersurface.

The next decompositions are motivated by previous works and ideas, clearly summarized in \cite{Gourgoulhon2012}. First, we introduce a time independent flat background metric $f_{ij}$, which coincides with $\gamma_{ij}$ at spatial infinity, and the following conformal decomposition:
\begin{equation}
\gamma_{ij}=\psi^{4}\ttg{ij}.
\label{eq:spmet}
\end{equation}
$\ttg{ij}$ is the conformal metric and the scalar $\psi:=(\gamma/f)^{1/12}$ is the conformal factor, where $\gamma=\det \gamma_{ij}$ and $f=\det f_{ij}$. $\tgg{ij}$ is defined by requiring that
\begin{equation}
\ttg{ik}\tgg{kj}=\delta_i^{\;j},
\end{equation}
which is equivalent to
\begin{equation}
\gamma^{ij}=\psi^{-4}\tgg{ij}.
\end{equation}
For asymptotically flat spacetimes, we have
\begin{equation}
N \to 1, \quad \beta^i \to  0, \quad \psi \to  1,
\end{equation}
at spatial infinity. Let us denote by $K^{ij}$ the extrinsic curvature on each hypersurface, given by
\begin{equation}
\begin{gathered}
K_{ij} = -\oneh \mathcal{L}_{\bm{n}}\gamma_{ij}= 
-\oo{2N}\left( \partial_t \gamma_{ij} - D_i\beta_j-D_j\beta_i\right),
\end{gathered}
\end{equation}
where $\bm{n}$ is the unitary future-directed timelike vector normal to each hypersurface, $\mathcal{L}$ stands for the Lie derivative and $D$ is the Levi-Civita connection associated with $\gamma_{ij}$. We define the tensor $A^{ij}$ as its traceless part:
\begin{equation}
A^{ij}=K^{ij}-\frac{1}{3}K\gamma^{ij},
\end{equation}
where $K$ represents the trace of the extrinsic curvature, $K=\gamma_{ij} K^{ij}$. We also define $h^{ij}=\tgg{ij}-f^{ij}$; the components of this tensor are not necessarily small, since we are not restricting this spacetime metric to perturbation regimes with respect to the flat spacetime. For asymptotically flat spacetimes we have
\begin{equation}
h^{ij} \to 0, \quad A^{ij} \to  0,
\end{equation}
at spatial infinity. Moreover, the gauge freedom of Einstein equations allows us to impose 4 additional conditions. In the case of the FCF, these conditions are $K=0$ and 
\begin{equation}
\D_k\tgg{ki}=0,
\label{eq:DiracG}
\end{equation}
where $\D$ is the Levi-Civita connection associated with $f^{ij}$. $K=0$ is the so-called maximal slicing condition, and Eq.~\eqref{eq:DiracG} is the generalized Dirac gauge. The next step is introducing a conformal decomposition of the extrinsic curvature,
\begin{equation}
K^{ij}=\psi^{-10} \hat A^{ij},
\end{equation}
and its longitudinal/transverse decomposition,
\begin{equation}
\hat{A}^{ij}=(LX)^{ij}+\hat{A}^{ij}_{TT},
\label{eq:hatA}
\end{equation}
where we have a vector field $X^i$ and the traceless and transverse tensor $\hat A^{ij}_{TT}$ satisfies $\D_i\hat{A}^{ij}_{TT}=0$.  $L$ is the conformal Killing operator defined as
\begin{equation}
(LX)^{ij}:=\D^iX^j+\D^jX^i -\frac{2}{3}f^{ij}\D_k X^k.
\end{equation}
The condition $\D_i\hat{A}^{ij}_{TT}=0$, up to boundary conditions, determines $X^i$. At spatial infinity, we will consider $\hat{A}^{ij}_{TT}\to 0$. These last two decompositions are motivated by the resolution of local uniqueness issues in the elliptic sector, as described in \cite{cordero2009improved}.

The following projections of the energy-momentum tensor are introduced for completeness:
\begin{gather*}
S_{ij}=T_{\mu\nu}\gamma^\mu_i\gamma^\nu_j, \\
S^i=-\gamma^{i\mu}T_{\mu\nu}n^{\nu}, \\
S=\gamma^{ij}S_{ij},\\
E=T_{\mu\nu}n^{\mu}n^{\nu},
\end{gather*}
and also the rescaled quantities $S^*_i=\psi^6S_i$, $S^*=\psi^6S$ and $E^*=\psi^6 E$. After all these definitions, we end up with two evolution equations for $h^{ij}$ and $\hat{A}^{ij}$,
\begin{eqnarray}	
\partial_t h^{ij} && = \beta^k\mathcal{D}_kh^{ij}-\tgg{ik}\mathcal D_k\beta^j-\tgg{kj}\mathcal D_k\beta^i+\frac{2}{3}\tgg{ij}\mathcal D_k\beta^k \nonumber \\ &&+2N\psi^{-6}\hat A^{ij}_{TT}+2N\psi^{-6}(LX)^{ij} ,\label{eq:hold}\\
 \partial_t \hat{A}^{ij}&&=S_{\hat{A}}^{ij},
 \label{eq:hold2}
\end{eqnarray}
and four elliptic equations for $X^i$, $\psi$, $N$ and $\beta^i$ (or combinations of these quantities). The starred variables are introduced because they are directly associated with the conserved quantities evolved in the fluid equations. The lengthy right hand side of Eq.~\eqref{eq:hold2} is omitted, but its explicit form can be found in \cite{cordero2009improved}.

\section{\label{sec:ref}Reformulation of the equations}

Many of the manipulations in this section have in mind the PNEs of the variables we are considering.
This means an expansion of the variables in powers of $1/c$, valid in the approximation of weak gravity and small velocities of the sources. Post-Newtonian hydrodynamics has been considered by a number of authors in the past \citep{blanchet1990post,Asada1996,Barausse2013}. To the leading order, the PNE of the metric variables we are manipulating can be deduced from \cite{blanchet1990post}, and are listed below:
\begin{equation}
\begin{gathered}	
N=1-\frac{U}{c^2}+\OO\left(\oo{c^4}\right), \;
\psi=1+\frac{U}{2c^2}+\OO\ooa 4, \\
N\psi^2=1+\OO\ooa 4, \\
\beta^i=\OO\left(\frac{1}{c^3}\right), \\
h^{ij}=\oot 4,  \\
\hat A^{ij}=\oot 3, \; X^i=\oot 3, \\
\hat A^{ij}_{TT}=\oot 5,
\end{gathered}
\label{eq:pNorders}
\end{equation}
where $U$ is the Newtonian potential. 

In light of the previous considerations regarding post-Newtonian orders~\eqref{eq:pNorders}, we reformulate the earlier version of the FCF formulation presented in~\cite{cordero2009improved}, aiming to improve numerical accuracy. 

For example, Equation~\eqref{eq:hold} motivates the introduction of a new variable involving $\beta^i$ and $X^i$, both of order $\OO(c^{-3})$. $(L\beta)^{ij}$ is the leading term in the first source terms of Equation~\eqref{eq:hold}, while $2N\psi^{-6}(LX)^{ij}$ is the leading term in the last source terms of Equation~\eqref{eq:hold}. Accordingly, we introduce a new vector field with the goal of simplifying, at least partially, this equation:
\begin{equation}\label{eq:Vdef}
 V^i=2 N \psi^{-6} X^i - \beta^i.
\end{equation}
It turns out that the post-Newtonian order of this vector is $V^i=\OO(c^{-5})$, as it will be justified later in this section. Dealing with $V^i$ instead of $\beta^i$ we expect to solve the spacetime metric more accurately in those cases or regions where the post-Newtonian approximation is valid. Once $X^i$, $\psi$, $N$ and $V^i$ are known, the shift vector $\beta^i$ can be directly computed from the previous definition for $V^i$ (no need to solve the elliptical equation for $\beta^i$ anymore). The equation most positively affected by this new auxiliary variable is the evolution equation for the $h^{ij}$ tensor, which, in terms of the new variables, reads:
\begin{eqnarray}	
\partial_t h^{ij} && = \beta^k\mathcal{D}_kh^{ij}-h^{ik}\mathcal D_k\beta^j-h^{kj}\mathcal D_k\beta^i+\frac{2}{3}h^{ij}\mathcal D_k\beta^k \nonumber \\ &&+2N\psi^{-6}\hat A^{ij}_{TT}+(LV)^{ij} 
 - X^j\mathcal D^i(2N\psi^{-6})\nonumber \\ &&-X^i\mathcal D^j(2N\psi^{-6})+\frac{2}{3}f^{ij}X^k\mathcal D_k(2N\psi^{-6}).
\label{eq:h}
\end{eqnarray}
After the introduction of $V^i$, this equation has been slightly simplified and, more importantly, has balanced post-Newtonian orders in both sides of the equation, which are now both $\OO (c^{-5})$. The previous version of this equation, Eq.~\eqref{eq:hold}, had $\OO (c^{-5})$ at the left hand side, while the individual terms on the right hand side were $\OO (c^{-3})$. Of course, in the previous version, terms of order $\OO (c^{-3})$ must theoretically cancel out to become $\OO (c^{-5})$, but this may not always happen numerically. With the introduction of the new vector $V^i$, we avoid this potential numerical issue.

Let us now focus on the other evolution equation for $\hat{A}^{ij}$ in \cite{cordero2009improved}. Instead of $\hat{A}^{ij}$, we propose explicitly working with the variables $X^i$ and $\hat A^{ij}_{TT}$. Notice that $\hat{A}^{ij}$ is $\OO(c^{-3})$, while $\hat A^{ij}_{TT}$ is $\OO(c^{-5})$ in post-Newtonian order. Therefore, taking $\hat A^{ij}_{TT}$ as variable will solve the spacetime metric more accurately too when the post-Newtonian approximation is valid. In other cases, our reformulation is just a change of variables with no additional effects, beyond the slight reducing of the number of source terms in Eq.~\eqref{eq:h}. The resulting evolution equation for $\hat A^{ij}_{TT}$ reads:
\begin{widetext}
\begin{eqnarray}	
\partial_t \hat A^{ij}_{TT} && = 
\beta^k\D_k\hat A^{ij}-\hat A^{kj}\D_k\beta^i-\hat A^{ik}\D_k\beta^j+\frac{5}{3}\hat A^{ij}\D_k\beta^k +2N\psi^{-6}\tilde\gamma_{kl}\hat A^{ik}\hat A^{jl} + \frac{3}{4}N\psi^{-6}\tgg{ij}\ttg{lk}\ttg{nm}\hat A^{km}\hat A^{ln}\nonumber \\ &&
+ N\psi^2\tilde R^{ij}_* -\oo 4N\psi^2\tilde R\tgg{ij} + \oo 2(\tgg{kl}\D_kh^{ij}-\tgg{ik}\D_kh^{lj}-\tgg{kj}\D_kh^{il})\D_l(N\psi^2)\nonumber\\ &&
+4\psi^{-1}\tgg{ik}\tgg{jl}\D_k\psi\D_l(N\psi^2)+4\psi^{-1}\tgg{ik}\tgg{jl}\D_l\psi\D_k(N\psi^2)-2\psi^{-1}\tgg{ij}\tgg{kl}\D_k\psi\D_l(N\psi^2)\nonumber\\ &&
+\frac{N\psi^2}{2}\tilde\gamma^{kl}\D_k\left(\D_lh^{ij}\right)-8N\tgg{ik}\tgg{jl}\D_l\psi\D_k\psi+2N \tgg{ij}\tgg{kl}\D_k\psi\D_l\psi -\tgg{ik}\tgg{jl}\D_k\D_l(N\psi^2)	-(L\dot X)^{ij}\nonumber\\ &&
-8\pi N\psi^{10}S^{ij}+4\pi NS^*\tgg{ij},
\label{eq:A}
\end{eqnarray}
where we have introduced the new variable $\dot{X}^i=\partial_t X^i$ (for which we derive an elliptic equation below),
\begin{eqnarray}	
\tilde R^{ij}_*=\frac{1}{2}\left(-\D_lh^{ik}\D_kh^{jl}-\ttg{kl}\tgg{mn}\D_mh^{ik}\D_nh^{jl}+\ttg{nl}\D_kh^{mn}(\tgg{ik}\D_mh^{jl}+\tgg{jk}\D_mh^{il})\right)+\frac{1}{4}\tilde\gamma^{ik}\tgg{jl}\D_kh^{mn}\D_l\tilde\gamma_{mn},
\end{eqnarray}
\end{widetext}
and
\begin{equation}
\tilde R=\oo 4\tgg{kl}\D_kh^{mn}\D_l\ttg{mn}-\oo 2\tgg{kl}\D_kh^{mn}D_n\ttg{ml}.
\end{equation}
The left hand side of Eq.~\eqref{eq:A} has order $\OO(c^{-6})$, while the terms of the last two lines of the right hand side of this equation have order $\OO(c^{-4})$. This new expression significantly reduces the number of terms that should cancel out with respect to previous versions. Correcting the post-Newtonian order of these terms is a possibility that will be considered in the future. However, it will not be easy in general since it contains terms including matter variables. We note that for this equation there is no need to introduce specifically $V^i$ since it does not modify the post-Newtonian order of the different terms.

We are going to consider the elliptic equations in \cite{cordero2009improved}, with some modifications and simplifications due to the use of the vector field $V^i$, as well as an additional elliptic equation for the new variable $V^i$ (instead of the elliptic equation for the shift vector $\beta^i$), and another additional elliptic equation for the variable $\dot{X}^i$ to close the system. The whole elliptic sector is presented below, where the post-Newtonian order of the corresponding expressions is placed at the end of each equation, which is, at the same time, the order of the variable under resolution:
\begin{eqnarray}	
\Delta X^i && +\oo 3\D^i\D_jX^j =  -\tgg{im}\left(\D_k\ttg{ml}-\oo 2\D_m\ttg{kl}\right)\hat{A}^{kl} \nonumber \\ &&+ 8\pi\tgg{ij}(S^*)_j \hspace{0.0cm}=\oot 3;
\label{eq:X}
\end{eqnarray}
\begin{eqnarray}	
\tgg{kl}\D_k\D_l\psi &&= -2\pi\psi^{-1}E^*-\oo 8 \psi^{-7}\ttg{il}\ttg{jm}\hat A^{lm}\hat A^{ij} \nonumber \\ &&+\oo 8\psi\tilde R\hspace{0.0cm}=\oot 2;
\label{eq:psi}
\end{eqnarray}
\begin{widetext}
\begin{eqnarray}	
\tgg{ik}\D_i\D_k(N\psi^2) && = 2\psi^{-1}\tgg{ik}\D_k\psi\D_i(N\psi^2)-2\psi^{-2}(N\psi^2)\tgg{ik}\D_k\psi\D_i\psi + \frac{3}{4} \psi^{-8} (N\psi^2)\ttg{il}\ttg{jm}\hat A^{lm}\hat A^{ij} \nonumber \\ && 
+ \oo 4(N\psi^2)\tilde R+4\pi\psi^{-2} (N\psi^2)S^* \hspace{0.0cm}=\oot 4;
\label{eq:N}
\end{eqnarray}
\begin{eqnarray}	
\Delta V^i && +\frac{1}{3}\D^i\D_jV^j=
-h^{kj}\mathcal D_k\mathcal D_j V^i - \oo 3h^{ik}\mathcal D_k\mathcal D_j V^j
+2N\psi^{-6}\left(h^{kj}\mathcal D_k\mathcal D_j X^i+\frac{1}{3}h^{ik}\mathcal D_k\mathcal D_j X^j\right) + \tgg{kj}X^i\D_k\D_j(2N\psi^{-6}) \nonumber \\
&& +\oo 3\tgg{ik}X^j\D_k\D_j(2N\psi^{-6}) +\D_k(2N\psi^{-6})\left(2\tgg{kj}\D_jX^i+\oo 3\tgg{ik}\D_jX^j+\oo 3\tgg {ij}\D_jX^k-\hat A^{ik}\right)\hspace{0.0cm}=\oot 5.
\label{eq:V}
\end{eqnarray}
In the final equation~\eqref{eq:V}, we clearly observe (by referring to~\eqref{eq:pNorders}) what was previously stated regarding the post-Newtonian order of $V^i$. Notice also the improvement in numerical accuracy dealing with the variable $N\psi^2$, for which we have the elliptic equation \eqref{eq:N}, instead of $N\psi$ as it was proposed in \cite{cordero2009improved}. 

Finally, by taking the divergence of Equation~\eqref{eq:A}, we obtain an elliptic equation for the new variable $\dot{X}^i$:
\begin{eqnarray}	
\Delta \dot X^j && +\oo 3\D^j\D_i\dot X^i=
\beta^k\D_i\D_k\hat A^{ij} - \D_i\hat A^{ik}\D_k\beta^j - \hat A^{ik}\D_i\D_k\beta^j +\frac{2}{3}\hat A^{ij}\D_i\D_k\beta^k + \frac{5}{3}\D_i\hat A^{ij}\D_k\beta^k\nonumber \\ &&
 -\frac{1}{2}N\psi^{-6}\tgg{jl}\D_l\left(\ttg{in}\ttg{km}\hat A^{nm}\hat A^{ik}\right) -\psi^{-8}\tgg{jl}\ttg{in}\ttg{km}\hat A^{nm}\hat A^{ik}\D_l(N\psi^2)+8\psi^{-7}N\tgg{jl}\ttg{in}\ttg{km}\hat A^{nm}\hat A^{ik}\D_l\psi \nonumber \\ &&
+2N\psi^{-6}\D_i(\ttg{kl}\hat A^{ik}\hat A^{jl}) -16\psi^{-7}N\ttg{kl}\hat A^{ik}\hat A^{jl}\D_i\psi + 2\psi^{-8}\ttg{kl}\hat A^{ik}\hat A^{jl}\D_i(N\psi^2) \nonumber\\ &&
-\oo2\D_i(N\psi^2)\D_lh^{ik}\D_kh^{jl}
-\frac{1}{6}\tgg{kj}\D_kh^{il}\D_i\D_l(N\psi^2) 
-\tgg{ik}\D_ih^{jl}\D_k\D_l(N\psi^2) \nonumber \\ &&
- 8N\tgg{ik}\D_ih^{jl}\D_k\psi\D_l\psi+4N\tgg{jl}\D_lh^{ik}\D_i\psi\D_k\psi
+4\psi^{-1}\tgg{ik}\D_ih^{jl}\big(\D_l(N\psi^2)\D_k\psi+\D_k(N\psi^2)\D_l\psi\big)\nonumber \\ &&
-4\psi^{-1}\tgg{jl}\D_lh^{ik}\D_i(N\psi^2)\D_k\psi
+ \tilde R^{ij}_{**} \D_i(N\psi^2) + N\psi^2\D_i\tilde R^{ij}_{**}-\oo 2 N\psi^2\tgg{ij}\D_i\tilde R \nonumber \\ && -8\pi\psi^{-2}E^*\tgg{jl}\D_l(N\psi^2)+16\pi\psi^{-1}NE^*\tgg{jl}\D_l\psi
+16\pi\psi^{-1}NS^*\tgg{jl}D_l\psi \nonumber \\ &&
- 8\pi N\psi^{10}\D_iS^{ij}- 8\pi\psi^{8}S^{ij}\D_i(N\psi^2)-64\pi\psi^9NS^{ij}\D_i\psi \hspace{0.0cm}=\oot 4,
\label{eq:dX}
\end{eqnarray}
where
\begin{equation}
\tilde R^{ij}_{**}=\frac{1}{2}\left(-\ttg{kl}\tgg{mn}\D_mh^{ik}\D_nh^{jl}+\ttg{nl}\D_kh^{mn}(\tgg{ik}\D_mh^{jl}+\tgg{jk}\D_mh^{il})\right)+\frac{1}{4}\tilde\gamma^{ik}\tgg{jl}\D_kh^{mn}\D_l\tilde\gamma_{mn}.
\end{equation}
\end{widetext}

We can check that the hyperbolic and elliptic sectors can be solved hierarchically (as previously). Indeed, given the hydrodynamical variables $E^*,\,S^*,\,S^*_j,\,S^{ij}$, we have the hierarchical structure of Table \ref{tab:hier} where the equations include terms with progressively post-Newtonian orders. Note that $\tgg{ij}=f^{ij}+h^{ij}$, with $f^{ij}$ fixed and known background metric.

\begin{table}[htbp!]
\begin{center}
\begin{tabular}{ccc}
\hline\hline
 \thead{Variable under \\ resolution} & \thead{Computed \\from Equation} & \thead{PNE}  \\
 \hline \hline
 $h^{ij}$ & \eqref{eq:h} & $\OO(c^{-4})$ \\
  $\att{ij}$ & \eqref{eq:A} & $\OO(c^{-5})$ \\
$X^i(h^{ij}, \hat A^{ij}_{TT}, S^*_j)$ & \eqref{eq:X} & $\OO(c^{-3})$ \\
$\hat A^{ij}(X^i,\hat A^{ij}_{TT})$ & \eqref{eq:hatA} & $\OO(c^{-3})$ \\
$\psi(h^{ij}, \hat A^{ij}, E^*)$ & \eqref{eq:psi} & $1+\OO(c^{-2})$\\
$(N\psi^2)(h^{ij}, \hat A^{ij}, \psi, S^*)$ & \eqref{eq:N} & $1+\OO(c^{-4})$ \\
$N(\psi, N\psi^2)$ & -- & $1+\OO(c^{-2})$ \\
$V^i(h^{ij}, \hat A^{ij}, X^i, \psi, N)$ & \eqref{eq:V} & $\OO(c^{-5})$ \\
$\beta^i(X^i, \psi, N, V^i)$ & \eqref{eq:Vdef} &  $\OO(c^{-3})$ \\
$\dot X^i(h^{ij}, \hat A^{ij}, X^i, \psi, N, \beta^i, E^*, S^*, S^{ij})$ & \eqref{eq:dX} & $\OO(c^{-4})$ \\
\hline  
\end{tabular}
\end{center}
\caption{Hierarchical structure of resolution for the hyperbolic and elliptic sectors of the reformulated FCF.}
\label{tab:hier}
\end{table}

Note that in Eq.~\eqref{eq:N} we solve $N\psi^2$, with post-Newtonian order $\OO(c^{-4})$, instead of $N$ or $N\psi$, of order $\OO(c^{-2})$. However, there may be additional numerical errors due to the presence of the first derivate of $(N\psi^2)$ in the right hand side of Eq.~\eqref{eq:N}. This was not the case in the elliptic equation for $(N\psi)$ stated in \cite{cordero2009improved}. We will keep in mind potential numerical advantages when solving $(N\psi)$ instead of $(N\psi^2)$.

In the next Section we compute the stationary initial data associated with the spacetime geometry of a rotating neutron star as an illustrative numerical application of the proposed reformulation of the FCF.

\section{\label{sec:num}Computation of stationary spacetime initial data
}

\subsection{Initial set-up}
As a numerical test we use a rotating neutron star composed of a perfect fluid with polytropic equation of state $p = C\rho^\Gamma$, where $p$ is the pressure, $\Gamma = 2$ and $C = 145731$ (cgs units, $K=100$ in geometrized units). Considering a central density $\rho_c=7.91\cdot 10^{14}$ g/cm$^3$ and a rotation frequency $f = 550$ Hz, we obtain a neutron star with equatorial radius $R_{\text{eq}}=12.86$ km and polar radius $R_{\text{p}}=11.20$ km, and gravitational mass $M=1.487\,M_\odot$. Hydrodynamical and spacetime metric variables are computed with the code \verb|rotstar_dirac| \cite{ming2006rotating} based on the C++ library LORENE \cite{lorene}. This code employs the same gauge we use here, i.e., maximal slicing and Dirac gauge \eqref{eq:DiracG}. This allows us to use the hydrodynamical quantities to compute the spacetime metric variables with our reformulation. In addition, the spacetime metric variables from LORENE can be used to check the numerical results from our approach. For this compact object, spacetime is stationary and we can adapt the coordinate time $t$ to this stationarity, setting the derivatives with respect to $t$ in Eqs.~\eqref{eq:h} and \eqref{eq:A} to zero. Moreover, this spacetime is axisymmetric and we can also adapt spherical type coordinates $r$ and $\theta$ (2-dimensional problem). The rotation axis is fixed at $\theta =0$.

We are going to use a finite-difference grid, with equally spaced cells in the radial and angular directions. The numerical grid has $n_r=3200$ points in the radial coordinate $r$ and $n_{\theta}=64$ points in polar angle $\theta$, where
$$r_1 = \frac{\Delta r}{2}, \; r_{n_r}=12R_{\text{eq}}-\frac{\Delta r}{2}, \; \theta_1=\frac{\Delta\theta}{2}, \; \theta_{n_{\theta}}=\pi-\frac{\Delta\theta}{2},$$
being $\Delta r = 12R_{\text{eq}}/n_{r}\approx 48$ m and $\Delta\theta = \pi/n_{\theta}\approx 0.05$ the radial and angular cell sizes, respectively. This grid covers the spacetime up to $12$ times the radius of the star with more than $260$ radial points inside the star at the equator.

We use ghosts cells to compute the discretization of the spatial derivatives close to the numerical grid boundaries, keeping the same radial and angular cell sizes. The values of the numerical solution in the ghost cells for the different variables take into account the theoretical behaviour of these variables close to the boundaries. If
$u_{ij}$ denotes the numerical solution of the variable $u$, where $i$ and $j$ indexes refer to the radial and angular cells, respectively, then their values in the ghost cells are given by:
\begin{eqnarray}
 u_{(1-l)j}=\pm u_{lj}, \\
 u_{i(1-l)}=\pm u_{il}, \quad u_{i(n_\theta+l)}=\pm u_{i(n_\theta-l+1)},
\end{eqnarray}
where $l$ denotes the number of ghost cells in the corresponding direction, the positive sign denotes symmetric boundary conditions and the negative sign denotes antisymmetric boundary conditions. We are going to follow the symmetries used in \cite{baumgarte2013numerical}. Since we will need to solve some elliptic equations in orthonormal cylindrical components, we also need to consider symmetries in this new basis. Taking into account the transformations from spherical to cylindrical components, we follow the prescription of Table \ref{tab:symcyl}.

\begin{table}[htbp!]
\begin{center}
\begin{tabular}{lcccc}
\hline\hline
 & &  Center\hfill & Axis \hfill& Equator\hfill \\
 \hline
\multirow{3}{*}{Vectors $X^i$, $V^i$ and $\dot{X}^i$\hspace{0.5cm}} & $\rho$ & - & - & + \\
& $z$ & + & + & -\\
& $\varphi$ & - & - & +\\
\hline
\multirow{6}{*}{Tensor $h^{ij}$} & $\rho\rho$ & + & + & + \\
& $zz$ & + & + & +\\
& $\varphi\varphi$ & + & + & +\\
& $\rho z$ & - & - & -\\
& $\rho\varphi$ & + & + & +\\
& $z\varphi$ & - & - & -\\
\hline  \hline
\end{tabular}
\end{center}
\caption{Parity conditions for cylindrical orthonormal components of vectors and
tensors as implemented in the resolution of some elliptic equations.
Components of vectors $X^i$, $V^i$ and $\dot{X}^i$, and tensor $h^{ij}$ have to be multiplied with
the corresponding sign when they are copied into ghost zones.}
\label{tab:symcyl}
\end{table}

At the outer boundary we impose a Robin condition, assuming $u=u_0+M/r^n$ for a generic variable $u$, which is equivalent to impose $u\to u_0$ at spatial infinity ($r\to\infty$) and $\partial_r u = -n \,(u-u_0)/r$, so only the values $n$ and $u_0$ need to be specified. The values in the ghost cells of the outer boundary $u_{(n_r+l)j}$ are set such that this behaviour is reproduced. We consider $n=1$ for scalar fields and $n=2$ for vector fields. We will also need to solve an elliptic equation for the tensor $h^{ij}$ using orthonormal cylindrical components; in this case, $n=3$ will be used for the diagonal components and $n=4$ for the non-diagonal non-zero component $h^{\rho z}$. This is the case when we solve stationary initial data for a rotating neutron star in the Dirac gauge. In dynamical spacetimes, we must set $n=1$ for the $h^{ij}$ components in general.

All discretizations of the spatial differential operators are second-order and we use the LAPACK library to invert the Laplacian operators in the resolution of the elliptic equations. In some equations the variable under resolution appears outside the main Laplacian operator in the source term; in this case we apply a fix-point iterative method with a relaxation factor. Each loop finishes when the mean difference between two successive iterations is smaller than a given tolerance $\epsilon$,
$$\sum_{i=1}^{n_r}\sum_{j=1}^{n_{\theta}} \frac{|u^{k+1}_{ij}-u^{k}_{ij}|}{n_rn_{\theta}} < \epsilon,$$
where $k$ denotes the corresponding iteration and $(i,j)$ refers to a generic point of the numerical mesh. For all variables we use $\epsilon = 10^{-6}$.

\subsection{Vector and tensor Poisson-like equations}

In those cases where we solve a vector or tensor Poisson-like equation, we make a transformation to orthonormal cylindrical coordinates because there are more components decoupled with respect to the spherical coordinates. The vector Poisson-like equations we need to solve have the following form:
\begin{equation}
\Delta v^i + \lambda \, f^{ij}\D_j\D_kv^k = R^i.\label{eq:vtype}
\end{equation}
where $\Delta=\D_k\D^k$, $R^i$ is a general source term which does not depend on the unknown vector $v^i$ and $\lambda$ is a constant. The first step of the general procedure for solving Eq.~\eqref{eq:vtype} follows the strategy used in \cite{bonazzola2004fcf, grandclement2001multidomain}. It considers the divergence of the previous equation to derive
\begin{equation}
    \Delta \phi = \frac{\D_iR^i}{1+\lambda},
\end{equation}
with $\phi=\D_kv^k$. We solve this scalar Poisson equation and get $\phi$. The second step is to solve $v^i$ via the following equation (equivalent to the original one):
\begin{equation}
    \Delta v^i = R^i - \lambda \, f^{ij}\D_j\phi.
\end{equation}
In orthonormal cylindrical coordinates $\{\rho,\phi,z\}$ and in axisymmetry, all components decouple in the following way:
\begin{eqnarray}
(\Delta v)^{\rho} && = \Delta v^{\rho}-\frac{v^{\rho}}{\rho^2} = R^{\rho} - \lambda \, \D^{\rho}\phi, \\  
(\Delta v)^{\phi} && = \Delta v^{\phi}-\frac{v^{\phi}}{\rho^2} = R^{\phi}, \\ 
(\Delta v)^{z}  && = \Delta v^{z} = R^{z} - \lambda \, \D^{z}\phi.
\end{eqnarray}
The lef thand side of these equations is discretized applying second-order finite differences, leading to a linear operator, which we invert using LAPACK subroutines, as previously mentioned.

On the other hand, since we are imposing stationarity for the computation of stationary initial data, from Eq.~\eqref{eq:A}, we end up with a tensor Poisson-like equation for $h^{ij}$ of the following form:
\begin{equation}
(\Delta h)^{ij} = R^{ij},
\label{eq:htype}
\end{equation}
where $R^{ij}$ is a general source term that does not depend on the unknown tensor $h^{ij}$. In addition, in axisymmetry, we can consider coordinates in such a way that $h^{r\phi}=h^{\theta\phi}=0$. It turns out that in orthonormal cylindrical coordinates not all the non-zero components of $h^{ij}$ ($h^{\rho\rho}$, $h^{zz}$, $h^{\rho z}$ and $h^{\phi\phi}$) are fully decoupled, even if axisymmetry is imposed:
\begin{eqnarray}	
(\Delta h)^{\rho\rho} && = \Delta h^{\rho\rho}-\frac{2}{\rho^2}\left(h^{\rho\rho}-h^{\phi\phi}\right),\label{eq:hrr}\\  
(\Delta h)^{\phi\phi} && = \Delta h^{\phi\phi}+\frac{2}{\rho^2}\left(h^{\rho\rho}-h^{\phi\phi}\right), \label{eq:hpp}\\ 
(\Delta h)^{zz} && = \Delta h^{zz}, \label{eq:hzz}\\ 
(\Delta h)^{\rho z} && = \Delta h^{\rho z}-\frac{h^{\rho z}}{\rho^2}.\label{eq:hrz}
\end{eqnarray}
The equations for $h^{zz}$ and $h^{\rho z}$ are fully decoupled from the rest ones. If we define the auxiliary variables 
\begin{equation}	
C_1=h^{\rho\rho}+h^{\phi\phi},\;
C_2=h^{\rho\rho}-h^{\phi\phi},
\end{equation}
then we can derive the following decoupled elliptic equations:
\begin{eqnarray}	
\Delta C_1 &&= R^{\rho\rho}+R^{\phi\phi},\label{eq:C1}\\ 
\Delta C_2 -\frac{4}{\rho^2}C_2&&= R^{\rho\rho}-R^{\phi\phi}.\label{eq:C2} 
\end{eqnarray}
Once $C_1$ and $C_2$ are solved, we can easily get $h^{\rho\rho}$ and $h^{\phi\phi}$.

We will use this strategy to solve the whole system of elliptic equations.

\subsection{Comparison with results in the xCFC formulation}

\begin{figure}[htbp!]
    \includegraphics[width=\linewidth]{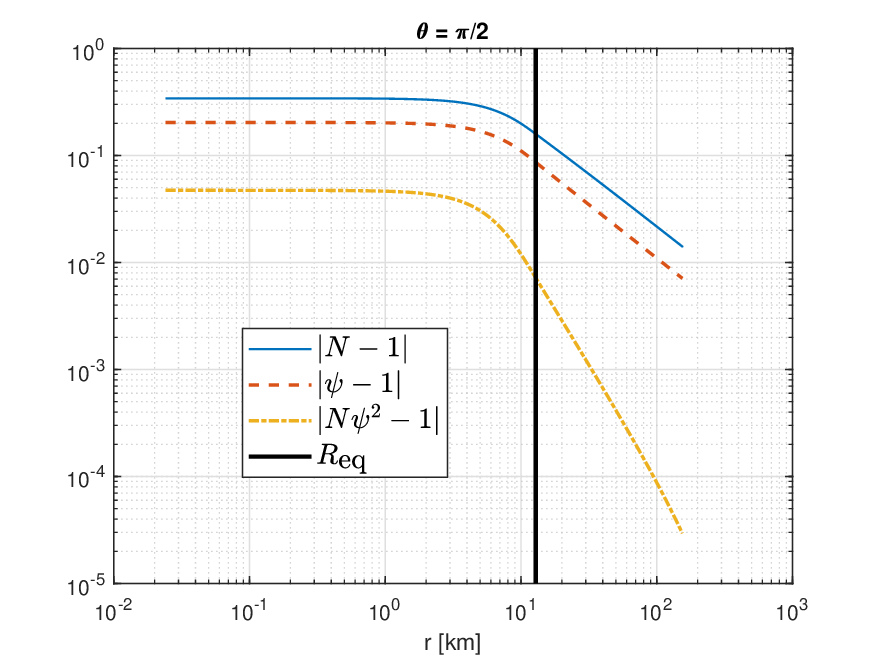}
\caption{Radial profiles of $(N-1)$ (solid blue line), $(\psi-1)$ (dashed red line), and $(N\psi^2-1)$ (dash-dotted yellow line) in log-log scale at $\theta=\pi/2$ are plotted. The vertical solid black line denotes the radius of the star at $\theta=\pi/2$.}
\label{im:Npsicomp3D}
\end{figure}

By xCFC formulation we mean imposing CFC condition with the decomposition considered in Eq.~\eqref{eq:hatA} and modifications presented in \cite{cordero2009improved} (where the auxiliary variables $V^i$ and $\dot{X}^i$ were not introduced). In this subsection we impose $h^{ij}=0$, so we can compare our results with the ones obtained with this xCFC formulation, that can be seen as an approximation to the FCF, except in the spherically symmetric case, where the xCFC formulation is exact (see \cite{cordero2011maximal}). Then, we will compare the numerical solution of Eqs.~\eqref{eq:X}--\eqref{eq:dX} imposing $h^{ij}=0$ to the resulting one using the original xCFC formulation. Indeed, it is convenient to solve an elliptic equation for $(N\psi^2-1)$ instead of $N\psi^2$, because $(N\psi^2-1)=\OO(c^{-4})$ in PNE. Therefore, the elliptic equations for $X^i$ and $\psi$ are the same as in the xCFC case, and the main differences in our proposal are: (i) we compute $N$ from the elliptic equation for $(N\psi^2 -1)$, derived from Eq.~\eqref{eq:N}; (ii) the shift vector is obtained once $V^i$ is known; and (iii) the elliptic equation for $\dot X^i$ was not considered in the original xCFC formulation.
In Figure \ref{im:Npsicomp3D} we check that the radial profile of $(N\psi^2-1)\sim\OO(c^{-4})$ decreases much faster than $(N-1)\sim\OO(c^{-2})$ and $(\psi-1)\sim\OO(c^{-2})$; for example, at $r=10^2$, $(N\psi^2-1)$ is 100 times smaller than the other two quantities.\\

\begin{figure}[htbp!]
    \includegraphics[width=\linewidth]{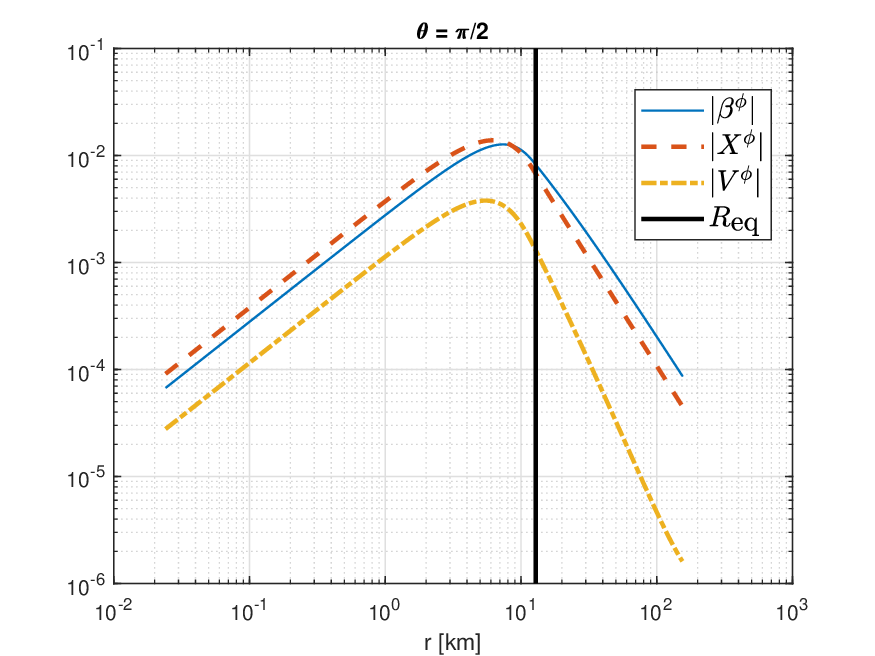}
\caption{Radial profiles of $\beta^i$ (solid blue line), $X^i$ (dashed red line), and $V^i$ (dash-dotted yellow line) in log-log scale at $\theta=\pi/2$. The solid vertical black line denote the radius of the star at $\theta=\pi/2$.}
\label{im:Vbcomp}
\end{figure}

\begin{figure}[htbp!]
    \includegraphics[width=\linewidth]{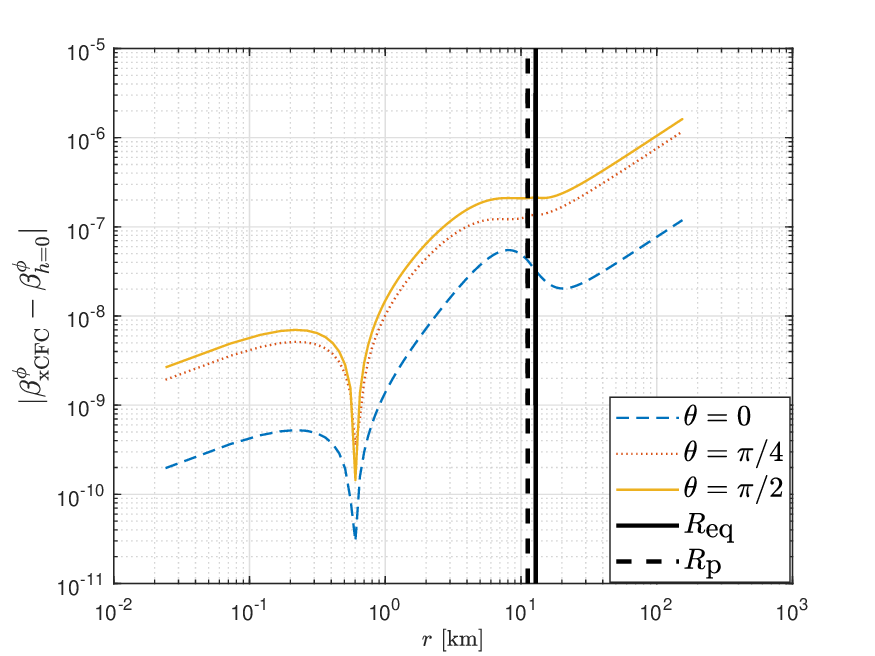}
\caption{Radial profiles of the difference between the computation of $\beta^\phi$ in the xCFC formulation, $\beta^\phi_{\text{xCFC}}$, and with our approach imposing $h=0$, $\beta_{h=0}$, in log-log scale at $\theta=0,\pi/4,\pi/2$. The solid and dashed vertical black lines denote the radius of the star at $\theta=\pi/2$ and $\theta=0$, respectively.}
\label{im:betacomp}
\end{figure}

Radial profiles of $V^i$ and the shift vector $\beta^i$ (directly computed from $V^i$) are displayed on Figure \ref{im:Vbcomp}.  
For these vector fields only the angular components $V^\phi$ and $\beta^\phi$ are non-zero. Let us remember that $V^i\sim\OO(c^{-5})$, while $X^i\sim\OO(c^{-3})$ and $\beta^i\sim\OO(c^{-3})$. We check in Figure \ref{im:Vbcomp} that $V^i$ decreases much faster than $X^i$ and $\beta^i$ getting to differences of almost two orders of magnitude at the outer boundary. When we compare $\beta^i$ computed through the xCFC formulation, $\beta_{\text{xCFC}}^i$, to that computed with our new approach, the difference is very small (i.e., around $10^{-6}$ at most) and increases with $r$, as it can be observed on Figure \ref{im:betacomp}, where the radial profiles of this difference in absolute value at $\theta=0,\pi/4,\pi/2$ are displayed. This small difference may be due to the fact that we are solving $V^i \sim \OO(c^{-5})$ (with an elliptic equation with source terms of the same post-Newtonian orders), and then we compute $\beta^i$ from $V^i$; this strategy may be more accurate for larger radial values with respect to the direct computation of $\beta_{\text{xCFC}}^i$.
Concerning the vector field $\dot{X}^i$, we know that it must be zero theoretically as $\partial_t$ is a Killing vector if the full metric is solved ($h^{ij}=0$ is not imposed). However, we can solve Eq.~\eqref{eq:dX} imposing $h^{ij}=0$ as an experiment; the expected error should be, at most, of the order of the tensor $h^{ij}=\OO(c^{-4})$ that we are neglecting. 
In fact, we obtain a maximum absolute value for $\dot{X}^i$ of order $10^{-5}$ km$^{-1}$ (relatively small compared to $M^{-1}\sim 0.5$~km$^{-1}$), which may come from the numerical treatment of the variables close to the origin $r=0$.

\subsection{Numerical resolution of stationary initial data in the FCF}
\label{subsec:FCF}
In this subsection we solve all the metric variables without the restriction $h^{ij}=0$. We obtain the numerical solution of Eqs.~\eqref{eq:X}--\eqref{eq:dX} (without imposing $h^{ij}=0$) for the variables $X^i$, $\psi$, $(N\psi^2-1)$ (at this point we can recover $N$ therefore), $V^i$ (at this point we can then recover $\beta^i$), and $\dot{X}^i$. Moreover, we solve Eqs.~\eqref{eq:h} and \eqref{eq:A} setting to zero the time derivatives, as we are computing a stationary spacetime. See more details in Table \ref{tab:hier}, already presented at the end of the previous section. On one hand, from Eq.~\eqref{eq:A}, we get an elliptic equation for $h^{ij}$:
\begin{widetext}
\begin{eqnarray}	
\tilde\gamma^{kl}\D_k\D_lh^{ij}&& = -\left(\frac{N\psi^2}{2}\right)^{-1}\Big(
\beta^k\D_k\hat A^{ij}-\hat A^{kj}\D_k\beta^i-\hat A^{ik}\D_k\beta^j+\frac{5}{3}\hat A^{ij}\D_k\beta^k +2N\psi^{-6}\tilde\gamma_{kl}\hat A^{ik}\hat A^{jl} \nonumber \\ &&+ \frac{3}{4}N\psi^{-6}\tgg{ij}\ttg{lk}\ttg{nm}\hat A^{km}\hat A^{ln}
+ N\psi^2\tilde R^{ij}_* -\oo 4N\psi^2\tilde R\tgg{ij} + \oo 2(\tgg{kl}\D_kh^{ij}-\tgg{ik}\D_kh^{lj}-\tgg{kj}\D_kh^{il})\D_l(N\psi^2)\nonumber\\ &&
\left.
+4\psi^{-1}\tgg{ik}\tgg{jl}\D_k\psi\D_l(N\psi^2)+4\psi^{-1}\tgg{ik}\tgg{jl}\D_l\psi\D_k(N\psi^2)-2\psi^{-1}\tgg{ij}\tgg{kl}\D_k\psi\D_l(N\psi^2)\right.\nonumber\\ && \left.
-8N\tgg{ik}\tgg{jl}\D_l\psi\D_k\psi+2N \tgg{ij}\tgg{kl}\D_k\psi\D_l\psi -\tgg{ik}\tgg{jl}\D_k\D_l(N\psi^2)	-(L\dot X)^{ij}\right.\nonumber\\ &&
 -8\pi N\psi^{10}S^{ij}+4\pi NS^*\tgg{ij}\Big),
\label{eq:hell}
\end{eqnarray}
and, on the other hand, from Eq.~\eqref{eq:h}, an algebraic equation for $\hat A_{TT}^{ij}$:
\begin{eqnarray}	
\hat A^{ij}_{TT} && = -(2N\psi^{-6})^{-1}\left(\beta^k\mathcal{D}_kh^{ij}-h^{ik}\mathcal D_k\beta^j-h^{kj}\mathcal D_k\beta^i+\frac{2}{3}h^{ij}\mathcal D_k\beta^k+(LV)^{ij} \right.\nonumber \\ && 
\left.- X^j\mathcal D^i(2N\psi^{-6})-X^i\mathcal D^j(2N\psi^{-6})+\frac{2}{3}f^{ij}X^k\mathcal D_k(2N\psi^{-6})\right).
\label{eq:Aalg}
\end{eqnarray}
\end{widetext}
The term $(L\dot{X})^{ij}$ in \eqref{eq:hell} can be neglected since we are computing stationary initial data. We can also check the accuracy of our approach by including this vector field, solving it via its elliptic equation and monitoring the values of this quantity. We will discuss the results of both options at the end of this subsection. Eq.~\eqref{eq:h} already had source terms with consistent post-Newtonian order, so Eq.~\eqref{eq:Aalg} too. Moreover, neglecting the term $\partial_t \hat A^{ij}_{TT}$ in Eq.~\eqref{eq:A} makes the elliptic equation~\eqref{eq:hell} to also have a consistent post-Newtonian truncation in the source terms for the computation of the $h^{ij}$ tensor.\\

\begin{figure}[h]
\centering
 \includegraphics[width=\linewidth]{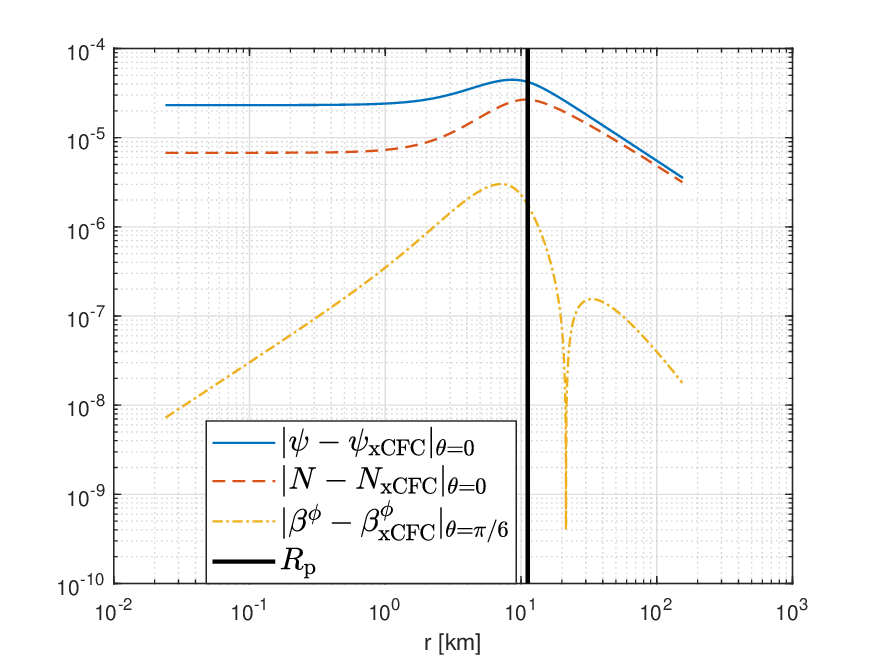}
\caption{Radial profiles of the difference between the corresponding variable obtained using the xCFC and the FCF formulations, in absolute value, for the conformal factor $\psi$, lapse $N$ and angular component of the shift $\beta^\phi$, at the polar angle $\theta$ where a maximum value is reached. The location of surface of the star at $\theta=0$ is denoted with a solid black lines.}
\label{im:comp3D}
\end{figure}

We overview here the iterative strategy used to solve all equations to obtain the stationary spacetime metric initial data, based on the hierarchical structure of the equations already mentioned in Table \ref{tab:hier}:
\begin{itemize}
    \item[1.] Solve Eqs.~\eqref{eq:X}--\eqref{eq:dX} for $X^i$, $\psi$, $(N\psi^2-1)$ ($N$ can therefore be also computed), $V^i$ ($\beta^i$ can therefore be also computed) and $\dot X^i$, imposing $h^{ij}=0$.
    \item[2.] Solve the elliptic equation \eqref{eq:hell} for $h^{ij}$, using the values computed in the previous step.
    \item[3.] Compute $\hat A_{TT}^{ij}$ from Eq.~\eqref{eq:Aalg}.
    \item[4.] Solve Eq.~\eqref{eq:X} for $X^i$, now considering the values computed previously (in general, $h^{ij}\neq 0$).
    \item[5.] Calculate $\hat A^{ij}$ from Eq.~\eqref{eq:hatA} (in general, $h^{ij} \neq 0 \neq \hat A_{TT}^{ij}$).
    \item[6.] Solve Eqs.~\eqref{eq:psi}--\eqref{eq:dX} for $\psi$, $(N\psi^2-1)$ ($N$ can be therefore also computed), $V^i$ ($\beta^i$ can therefore be computed), and $\dot X^i$ (in general, $h^{ij}\neq 0$).
    \item[7.] Go to Step 2, until a desired level of tolerance is achieved.
\end{itemize}

After only 5 iterations with this strategy, we observe that the absolute values of the differences of the variables in successive iterations are smaller than 1\%.\\

\begin{figure}[htbp!]
\begin{subfigure}{0.87\linewidth}
    \includegraphics[width=\linewidth]{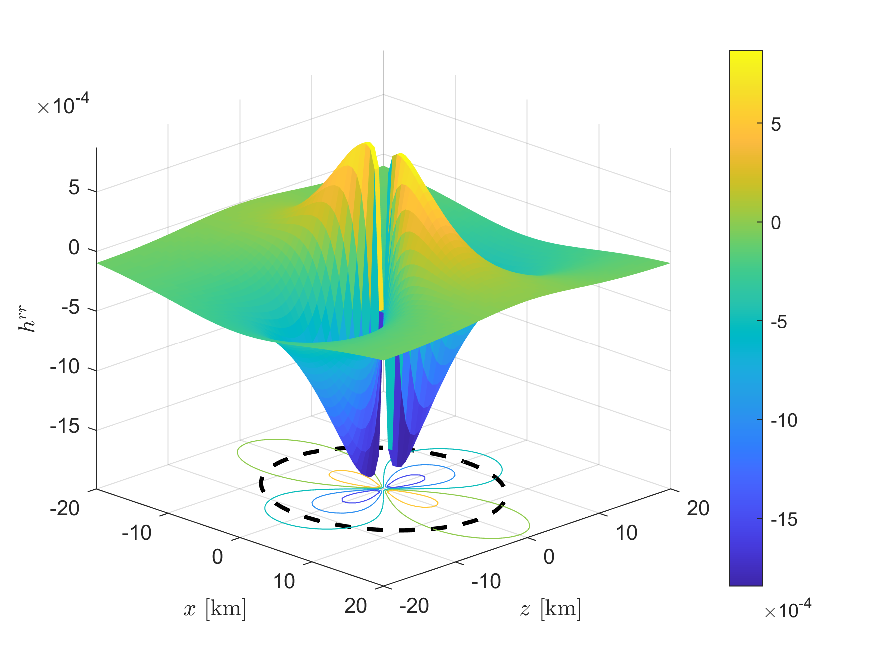}
    \end{subfigure}
 \\
\begin{subfigure}{0.87\linewidth}
    \includegraphics[width=\linewidth]{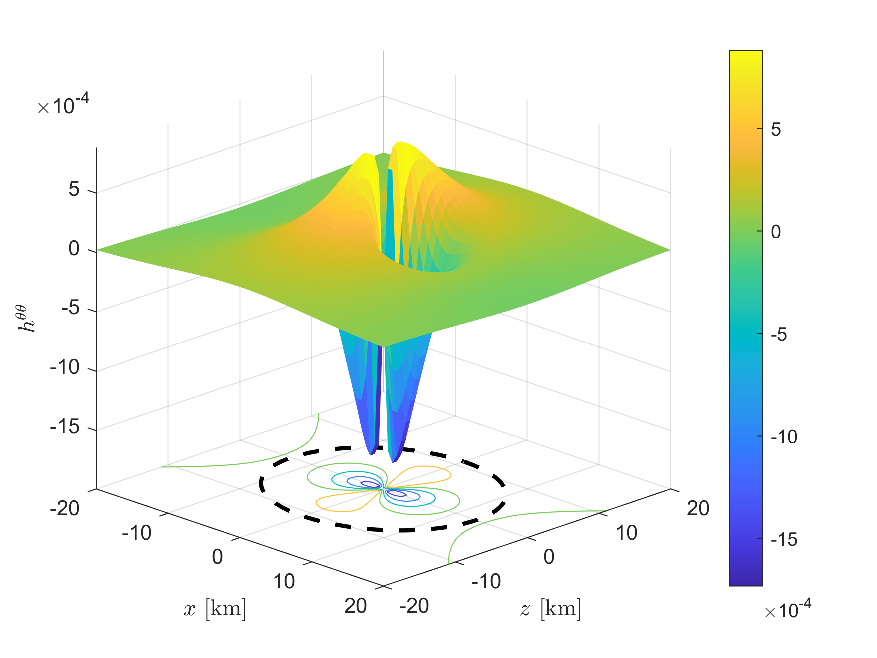}
    \end{subfigure}
\\
\begin{subfigure}{0.87\linewidth}
    \includegraphics[width=\linewidth]{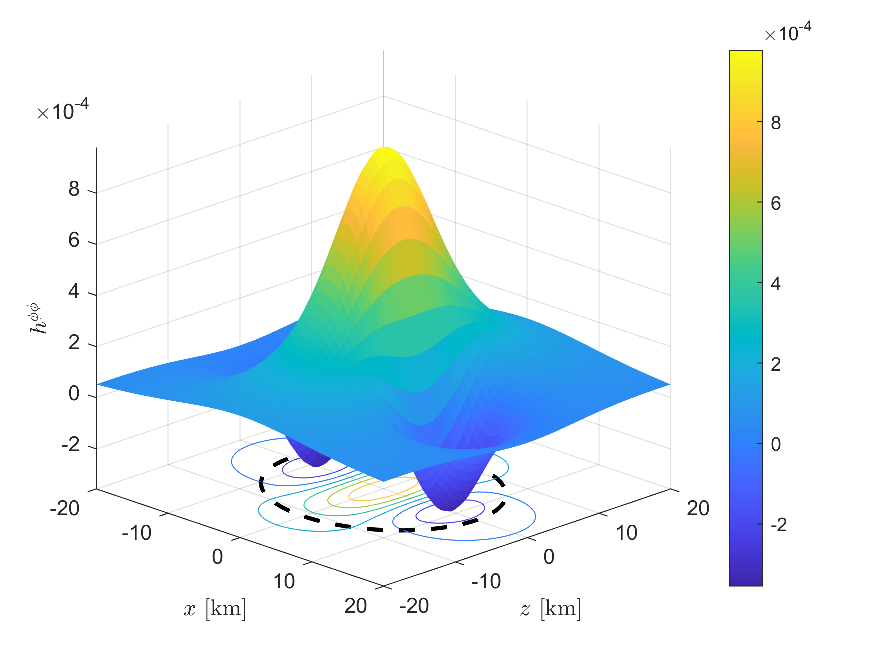}
    \end{subfigure}
 \\
\begin{subfigure}{0.87\linewidth}
    \includegraphics[width=\linewidth]{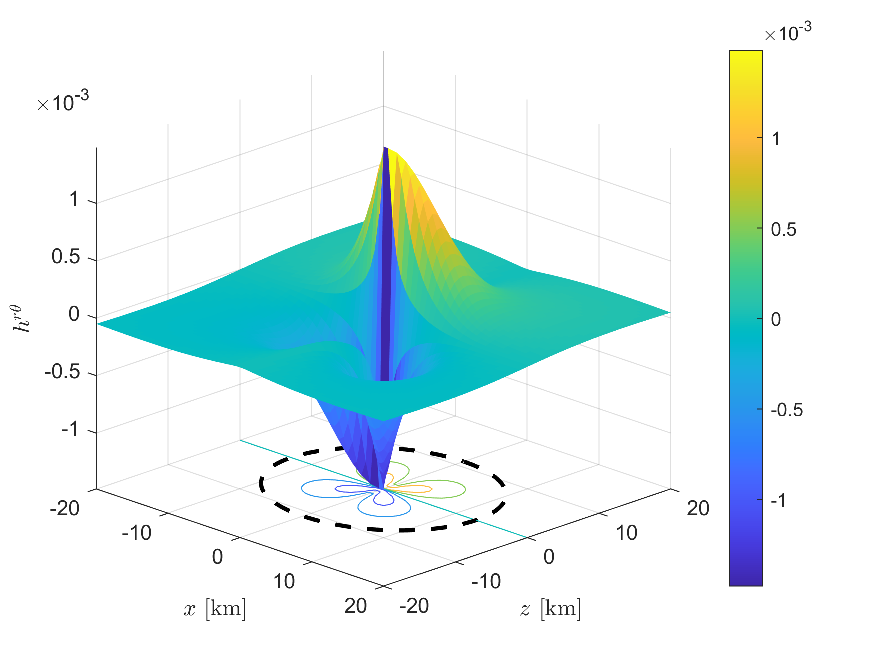}
    \end{subfigure}
\caption{Profiles of the non-zero components of $h^{ij}$ in the meridional plane. The location of surface of the star is denoted with a dashed black line. The rotation axis is placed at $x=0$.}
\label{im:hij3D}
\end{figure}

We obtain similar profiles for all variables with respect to the results shown in the previous subsection. A detailed analysis of the expected difference between the variables' values when $h^{ij}=0$ (xCFC) or $h^{ij}\neq 0$ (FCF) are considered can be found in the Appendix of \citep{cordero2009improved}. There, it is established that these differences are expected to be of the order of $h^{ij}$. We can check this fact numerically in Figure \ref{im:comp3D}, where we show the differences of $N$, $\psi$, and $\beta^i$ when cases $h^{ij}\neq 0$ and $h^{ij}=0$ are considered. We notice that these differences are several orders of magnitude smaller than the corresponding variables, and also smaller than the profiles of the non-zero components of $h^{ij}$ (when $h^{ij}\neq 0$ is considered) shown in Figure \ref{im:hij3D}.
\begin{figure}[htbp!]
\centering
    \includegraphics[width=0.99\linewidth]{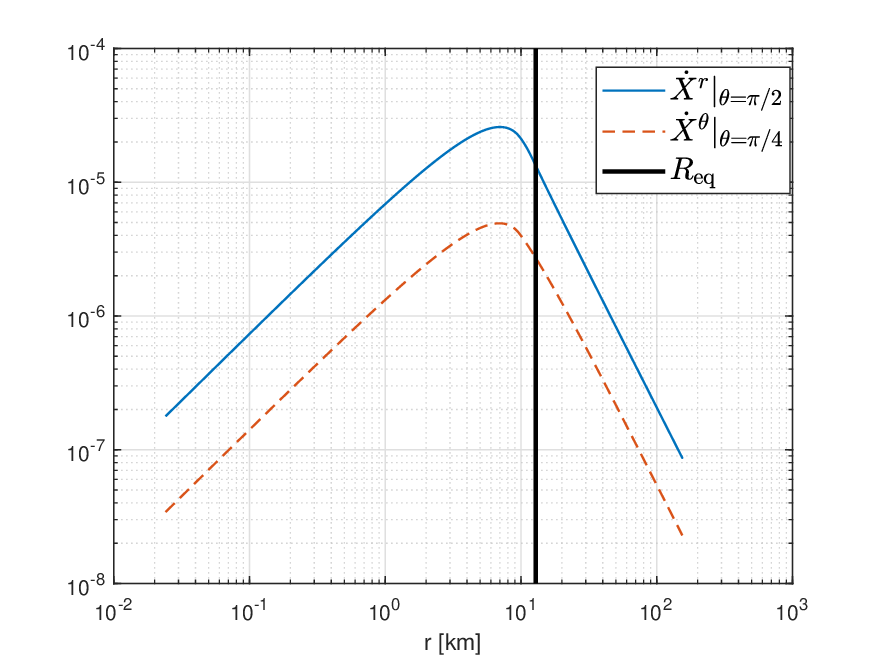}
\caption{Radial profiles of $\dot X^{r}$ at $\theta=\pi/2$ and $\dot X^{\theta}$ at $\theta=\pi/4$, where maximum values are reached.}
\label{im:dX3D}
\end{figure}
\begin{figure}[htbp!]
\centering
\begin{subfigure}{0.85\linewidth}
    \includegraphics[width=\linewidth]{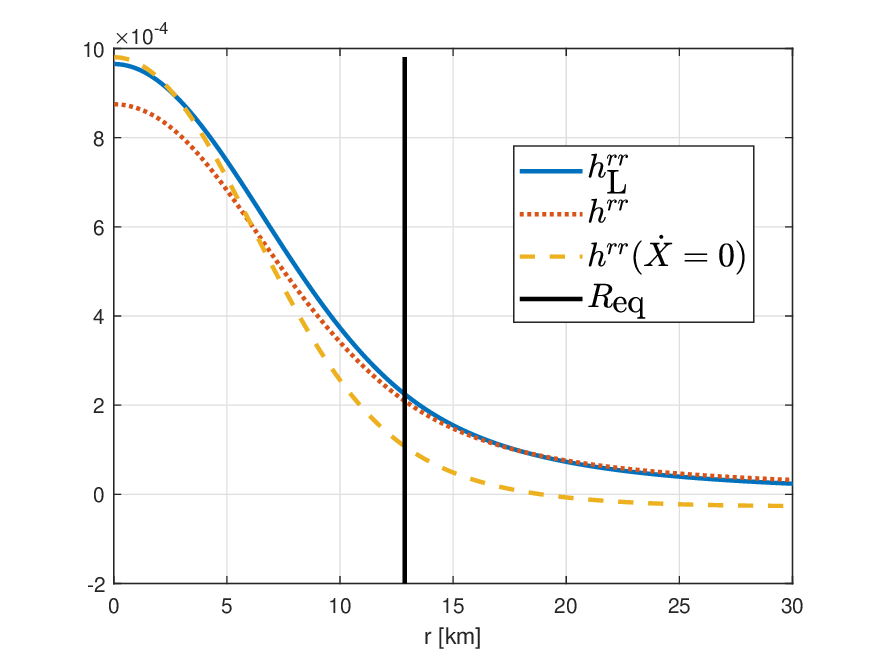}
    \end{subfigure}
 \\
    \begin{subfigure}{0.85\linewidth}
    \includegraphics[width=\linewidth]{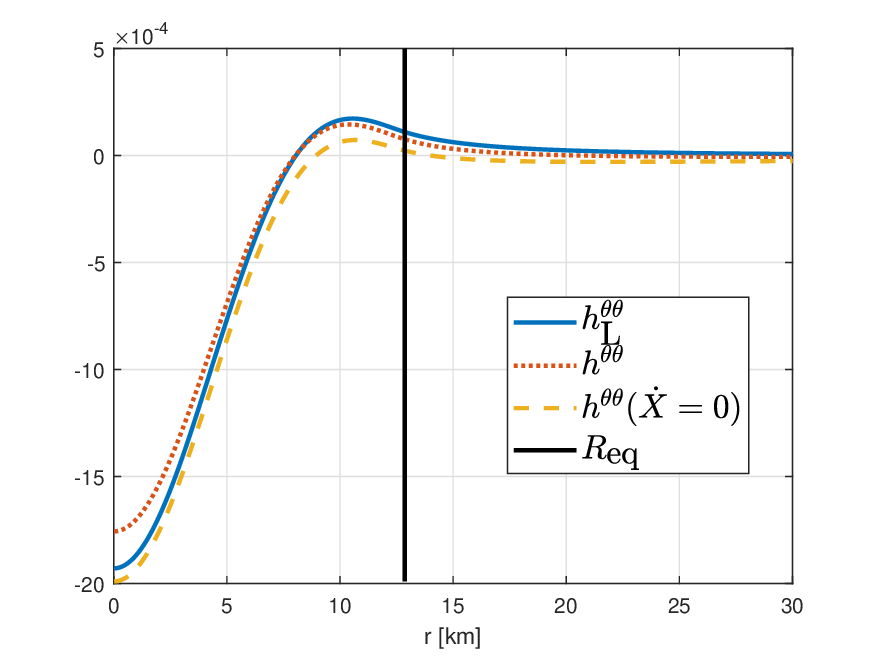}
    \end{subfigure}
    \\
    \begin{subfigure}{0.85\linewidth}
    \includegraphics[width=\linewidth]{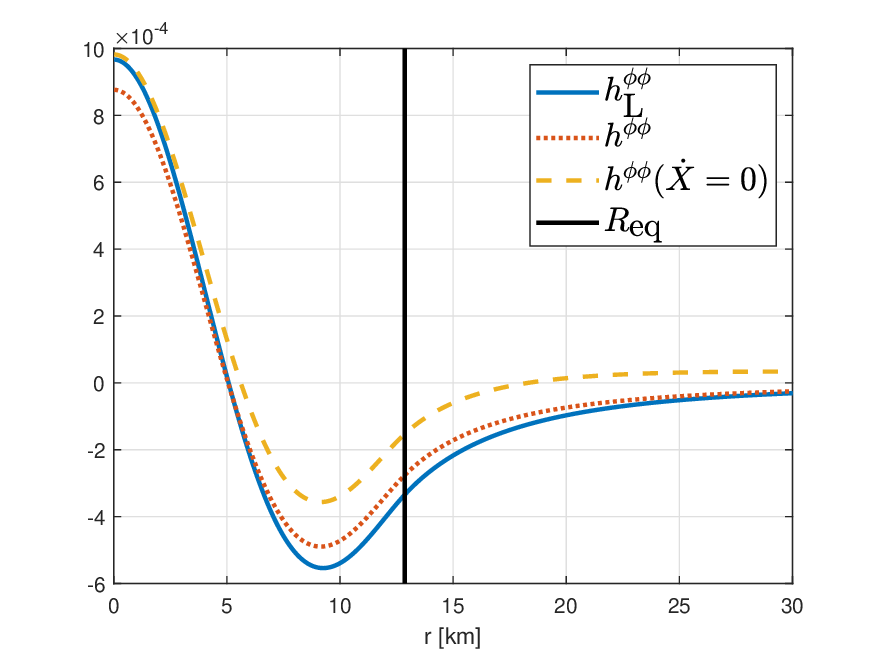}
    \end{subfigure}
\\
    \begin{subfigure}{0.85\linewidth}
    \includegraphics[width=\linewidth]{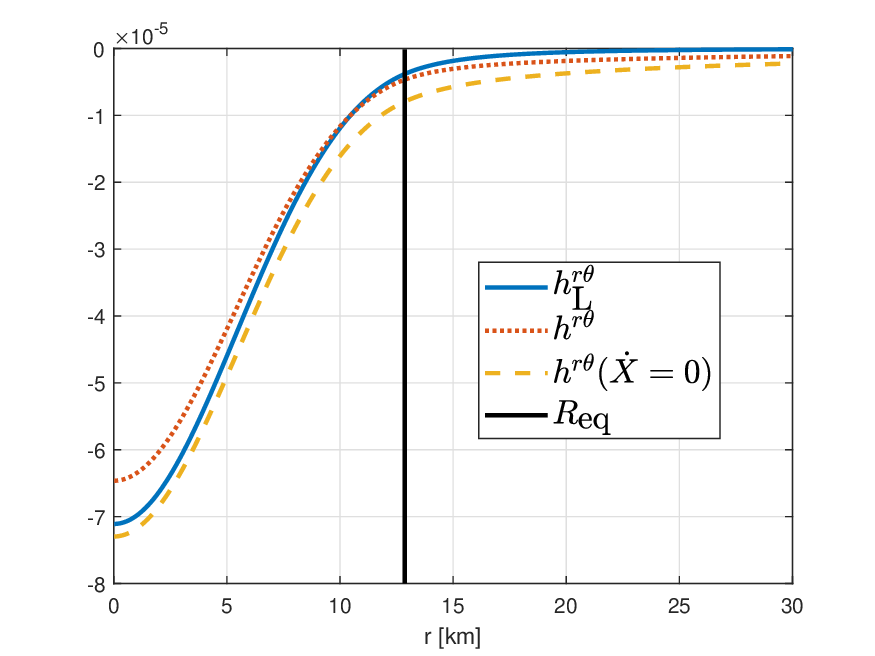}
    \end{subfigure}
\caption{Radial profiles at $\theta=\pi/2$ of the numerical solution for (from top to bottom) $h^{rr}$, $h^{\theta\theta}$, $h^{\phi\phi}$ and $h^{r\theta}$ imposing $\dot X^i=0$ (dashed yellow lines) and solving the $\dot X^i$ vector field (red dotted lines). The reference solution computed with LORENE is also displayed (solid blue lines). $\Delta r=96$ m is used.}
\label{im:comp}
\end{figure}

We get similar profiles for $\dot{X}^i$ in comparison with those computed in the previous subsection considering $h^{ij}=0$. The results for $\dot{X}^i$ are shown in Figure \ref{im:dX3D}. Fourth-order interpolation near $r=0$ for some second derivatives in the source of Eq.~\eqref{eq:dX} was required to avoid some numerical divergences at the center $r=0$ (possibly due to the appearance of some numerical errors and their amplification due to the use of orthonormal spherical components). It is interesting to observe that, on one hand, if we neglect the vector $\dot{X}^i$ in general (and in particular the term involving this vector in Eq.~\eqref{eq:hell}), we do not need anymore to apply the previously mentioned interpolation and we obtain profiles of $h^{ij}$ that are closer to the reference solution obtained with LORENE close to the center $r=0$. Recall that LORENE considers $\dot X^i=0$ to compute the initial data. On the other hand, we get profiles closer to those from the LORENE reference solution when $\dot{X}^i$ is included in the whole numerical domain except in the region close to the center. This means that interpolations to avoid numerical divergences are introducing also a source of error close to $r=0$. This behavior can be observed in Figure \ref{im:comp}, which will be further commented in the next subsection.

\begin{figure}[htbp!]
\begin{subfigure}{0.99\linewidth}
    \includegraphics[width=\linewidth]{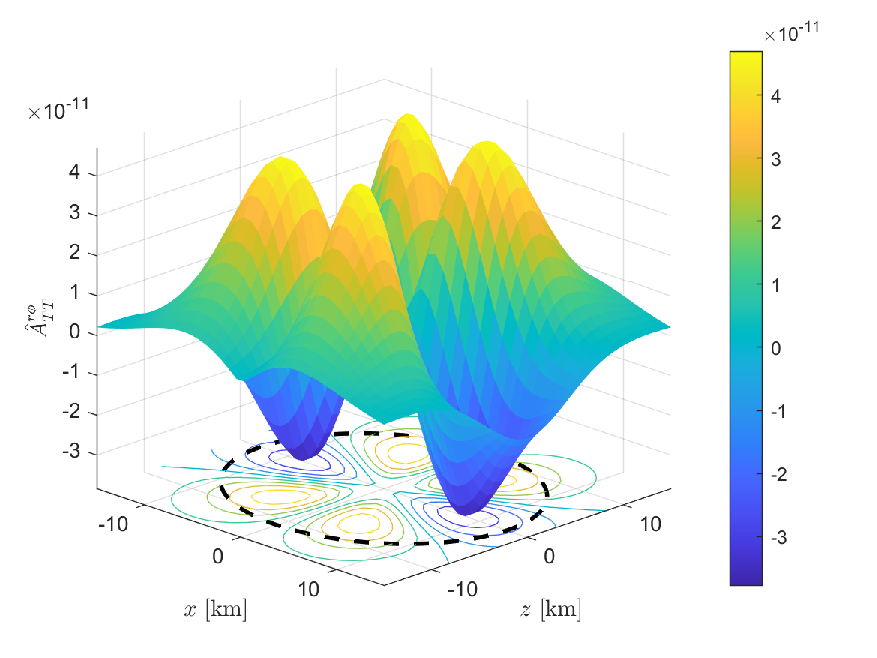}
    \end{subfigure}
    \\
    \begin{subfigure}{0.99\linewidth}
    \includegraphics[width=\linewidth]{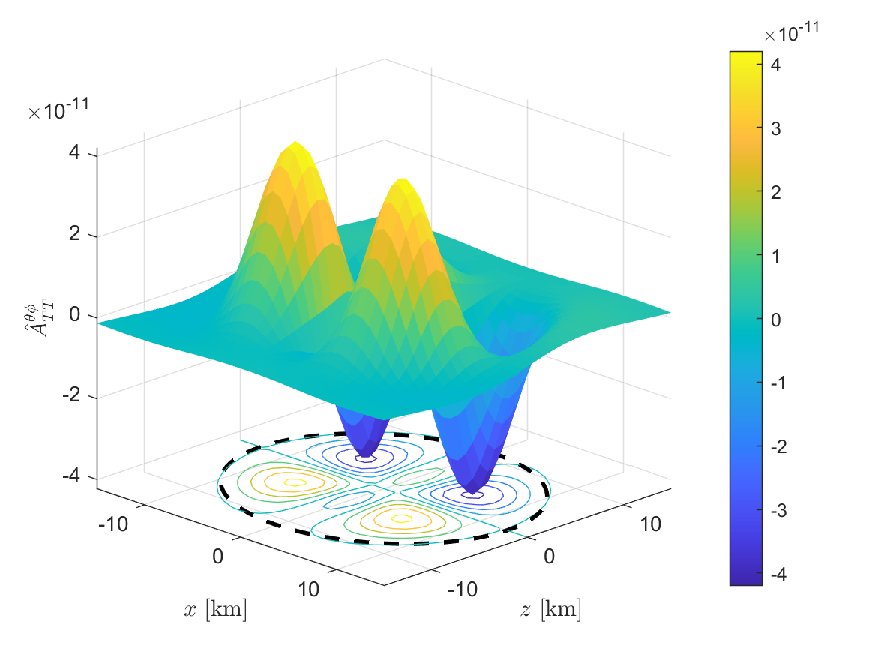}
    \end{subfigure}
\caption{Profiles of $\hat{A}^{r\phi}_{TT}$ (top) and $\hat{A}^{\theta\phi}_{TT}$ (bottom) in the meridional plane. A contour plot is also displayed. The location of surface of the star is denoted with dashed black lines. The rotation axis is placed at $x=0$.}
\label{im:ATT3D}
\end{figure}

Finally, in Figure \ref{im:ATT3D} we show the profiles for $\hat{A}^{ij}_{TT}$. We check that the components of $\hat{A}^{ij}_{TT}$ are at least two orders of magnitude smaller than the corresponding $\hat{A}^{ij}$ components, in agreement with their expected post-Newtonian orders. 

\subsection{Convergence}
We will compare with more detail our results with those computed with LORENE using several resolutions and placing the outer boundary at different radii. LORENE uses spectral methods and an exponential convergence with increasing resolution for smooth variables is expected. In our case, we use finite differences of second-order, so our convergence with increasing resolution is expected to be slower. Then, we use LORENE solutions as reference for testing the convergence and reliability of our approach. We compute the relative error between our numerical solution $f$ and the one derived with LORENE $f_L$, $\varepsilon_r$, using the discrete 2-norm.

We set a decay of the form $r^{-n}$ at the outer boundary and expect better results as the outer boundary is placed further from the neutron star. We will focus on the $h^{ij}$ tensor in this subsection, since the rest of the variables are not computed with LORENE (these variables were just introduced in this work) or we see no significant difference when increasing the resolution or changing the location of the outer boundary. The latter point may be due to other more dominant sources of numerical errors.

\begin{figure}[htbp!]
    \includegraphics[width=0.89\linewidth]{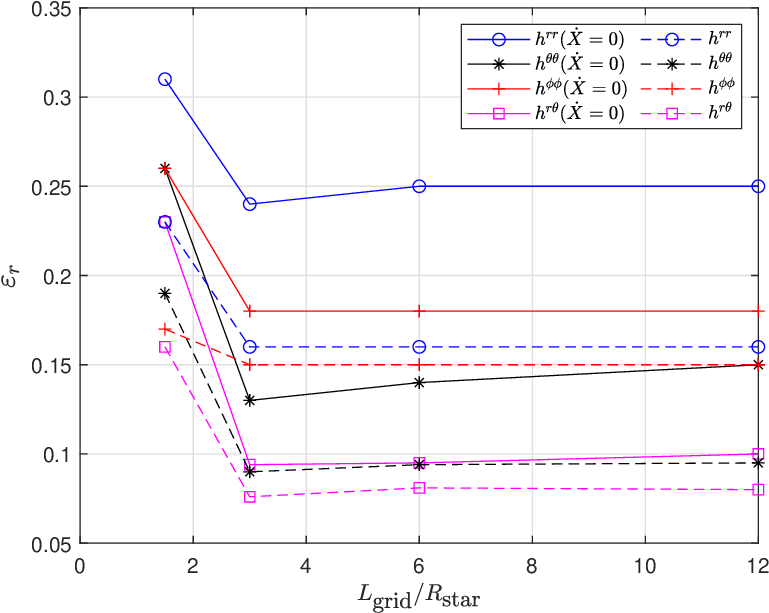}
\caption{Relative errors of the non-zero components of $h^{ij}$ having LORENE solutions as reference, both when $\dot X^i$ is set to zero (solid lines) and when is considered (dashed lines), for several locations of the outer boundary. $L_{\text{grid}}=k R_{\text{star}}$ with $k=1.5$ (blue), 3 (red), 6 (black), 12 (magenta). $n_r$ is such that the spatial resolution is fixed to $\Delta r=386$ m.}
\label{im:error_rel}
\end{figure}
We perform some simulations to establish the proper location of the outer boundary. In Figure \ref{im:error_rel} we show results for the relative errors of all the non-zero components of $h^{ij}$, including or not the computation of $\dot X^i$, respectively, with a fixed spatial resolution of $\Delta r=386$ m and varying the location of the outer boundary. The error stabilizes once the outer boundary is placed at 3 times the equatorial radius of the star $R_{\text{star}}$, or further away. If we are interested in dynamical spacetimes with potential gravitational radiation, the location of the outer boundary should be placed around 100 times the equatorial star radius, according to \cite{cordero2012gravitational}. 
From Figure \ref{im:error_rel}, we can quantify the consequences of neglecting $\dot X^i$. The errors are considerably higher if $\dot X^i=0$ is imposed. We think that improving the interpolation technique close to the center would reduce the relative errors even more when $\dot X$ is not neglected, although this is beyond the scope of this work.\\

Regarding the location of the outer boundary, we see a similar behavior with other tested resolutions: $\Delta r = 772$ m,  $193$ m, $96.4$ m and $48.2$ m. Since we concluded that placing the outer boundary at $r=3 R_{\text{star}}$ is enough for guaranteeing the numerical solution not to be strongly affected by the boundary condition, we fix the outer boundary at $r=6 R_{\text{star}}$.  All components stabilize its relative error for a spatial resolution of about $\Delta r \gtrsim 100$ m.\\

Notice that the relative errors between the numerical solution of the variables with our approach and using LORENE, shown in Figure \ref{im:error_rel}, are strongly affected by the values of the variables close to the outer boundary, which are several orders of magnitude smaller than those at the center. This fact can be shown in Figures \ref{im:comp}, where the non-zero components of $h^{ij}$ are displayed.\\

LORENE results are a very good reference solution. To conclude this subsection, we perform convergence studies with numerical results of our code considering different resolutions.

\begin{figure}[htbp!]
\centering
\includegraphics[width=\linewidth]{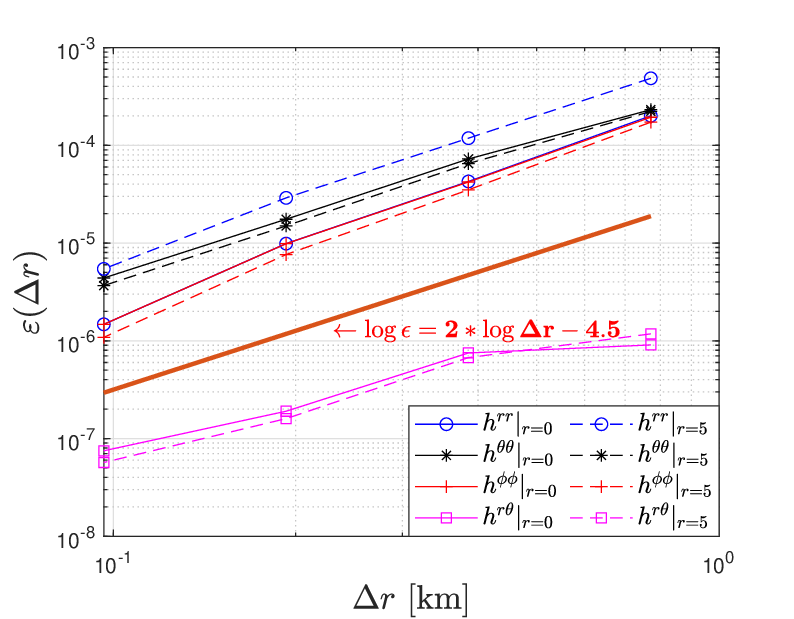}
\caption{$\varepsilon(\Delta r)$ at $\theta=\pi/2$ and $r=0$ km (solid lines) and $r=5$ km (dashed lines) for the non-zero components of $h^{ij}$ 
as a function of the numerical resolution. A bold red line has also been plotted     precisely marking a second-order behavior, as reference.}
\label{im:convours}
\end{figure}

We obtain in general an order of convergence close to the expected second-order, due to the use of second-order finite differences for the spatial derivatives. Check Figure \ref{im:convours}, where we plot the errors $\varepsilon (\Delta r) = |u(\Delta r)-u(\Delta r_h)|$, being $u(\Delta r)$ and $u(\Delta r_h)$ the numerical solutions with spatial resolutions $\Delta r$ and $\Delta r_h=48.2$ m, respectively.

\subsection{Dirac gauge and determinant condition.}

Dirac gauge condition is given by
\begin{equation}
    \mathcal D_i h^{ij} = 0.\label{eq:dh}
\end{equation}

In axisymmetry the condition $\mathcal D_i h^{i\phi}=0$ is trivially fulfilled. Although we assume the Dirac gauge to be satisfied, we are going to check it numerically, computing the following quantities:
\begin{equation}
    Q^i = \left.\frac{|\mathcal D_jh^{ji}|}{\max(|\mathcal D_{r}h^{ri}|,|\mathcal D_{\theta}h^{\theta i}|)}\right|_{ (r,\theta)=(r_0,\pi/2)},\;\; i = r,\theta,
\end{equation}
where $r_0$ stands for the radius where $|\mathcal D_jh^{ji}|$ has its maximum, i.e., the violation of the Dirac gauge condition is more significant. We choose $r_0=7$ km.\\

\begin{figure}[htbp!]
\begin{subfigure}{0.99\linewidth}
    \includegraphics[width=\linewidth]{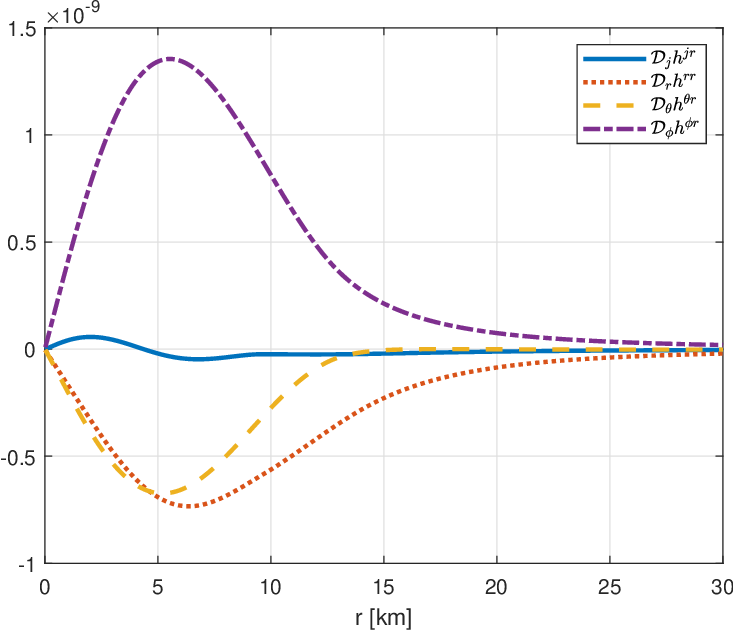}
    \end{subfigure}
 \\
    \begin{subfigure}{0.99\linewidth}
    \includegraphics[width=\linewidth]{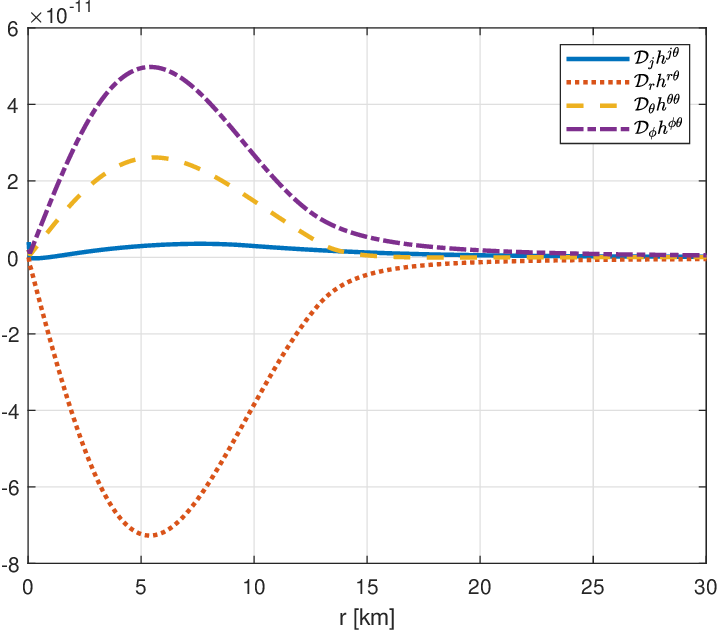}
    \end{subfigure}
\caption{Comparison of the gauge condition $\D_j h^{ji}$ (solid blue line) with respect to the individual covariant derivatives involved: $\D_rh^{ri}$ (dotted red line), $\D_\theta h^{\theta i}$ (dashed yellow line) and $\D_\phi h^{\phi i}$ (dash-dotted violet line) for $i=r$ (top subfigure) and $i=\theta$ (bottom subfigure).}
\label{im:dirac}
\end{figure}

The procedure used in LORENE to solve $h^{ij}$ \cite{ming2006rotating}, satisfies the Dirac gauge condition by construction. Of course, $Q^i$ is not going to be numerically zero, but one would expect to be much smaller than $1$. In fact, we get $Q^r_{\mbox{L}}, Q^\theta_{\mbox{L}}\sim 10^{-4}$, where the L subscript refers to quantities computed using LORENE data. Notice that our strategy does not impose the Dirac gauge condition directly (although it is applied when the expression in Eq. \eqref{eq:dh} appears explicitly). In our approach, we get $Q^r,Q^\theta\sim 10^{-2}$.
In Figure \ref{im:dirac} we compare the radial profiles of the Dirac gauge condition for $i=r$ and $i=\theta$ with respect to the individual covariant derivatives involved at $\theta=\pi/2$.\\

On the other hand, there is another condition that our metric must fulfill, according to Eq. \eqref{eq:spmet}:
\begin{equation}
    \det(\tilde{\gamma}^{ij}) = \det(f^{ij}+h^{ij})=f.
    \label{eq:det}
\end{equation}
In the case of the orthonormal spherical coordinates we are using in our numerical set up, $f=1$.\\

Reference \cite{ming2006rotating} also imposes this condition in the resolution of the $h^{ij}$ tensor, so the numerical solution derived with LORENE satisfies this algebraic constraint by construction. In our case, we compute the quantity $$\max(|1-\det(f^{ij}+h^{ij})|)$$ to check the violation of this algebraic constraint. We get a value of order $10^{-5}$, to be compared with zero. Since some variables are solved with a tolerance of $10^{-6}$, and taking into account that the absolute value of some components $|h^{ij}|$ reach values of order $10^{-3}$, we think that the constraint \eqref{eq:det} is reasonably satisfied.\\

Indeed, we think that a strategy combining the resolution of Eq. \eqref{eq:hell} with imposing somehow the Dirac gauge and determinant conditions of Eqs. \eqref{eq:dh} and \eqref{eq:det}, respectively, is quite important. We can always remove one differential equation for some component of the $h^{ij}$ and then apply the determinant condition to get this remaining component. The Dirac gauge in cylindric coordinates has only two non-zero components:
\begin{gather}
    \frac{\partial h^{\rho\rho}}{\partial\rho}+\frac{1}{\rho}(h^{\rho\rho}-h^{\phi\phi})+\frac{\partial h^{\rho z}}{\partial z} = 0,\label{eq:dirac1}\\
    \frac{\partial h^{\rho z}}{\partial \rho}+\frac{h^{\rho z}}{\rho}+\frac{\partial h^{zz}}{\partial z} = 0.\label{eq:dirac2}
\end{gather}

For instance, one can get first $h^{\rho z}$ from Eq. \eqref{eq:hrz}. Then, combining Eqs. \eqref{eq:hrr} and \eqref{eq:dirac1}, an elliptical equation for $h^{\rho\rho}$ can be derived:
\begin{equation}
    \Delta h^{\rho\rho}+\frac{2}{\rho}\frac{\partial h^{\rho\rho}}{\partial \rho} =S^{\rho\rho}-\frac{2}{\rho}\frac{\partial h^{\rho z}}{\partial z}.\label{eq:hrrd1}
\end{equation}
$h^{zz}$ can be derived from Eq. \eqref{eq:hzz}. Finally, $h^{\phi\phi}$ can be deduced from the determinant condition, or Eq. \eqref{eq:C1} for the scalar $C_1$ or Eq. \eqref{eq:C2} for the scalar $C_2$, as $h^{\rho\rho}$ is known. In this approach we only explicitly use the first equation of the Dirac gauge.\\

Another approach may be to get $h^{zz}$ from Eq. \eqref{eq:hzz} first. Then, the second Dirac gauge equation \eqref{eq:dirac2} and Eq. \eqref{eq:hrz} provide an elliptic equation for $h^{\rho z}$:
\begin{equation}
    \Delta h^{\rho z}-\frac{1}{\rho}\frac{\partial h^{\rho z}}{\partial \rho}-\frac{2}{\rho^2}h^{\rho z} =S^{\rho z}+\frac{1}{\rho}\frac{\partial h^{z z}}{\partial z}.\label{eq:hrzd2}
\end{equation}
Now, if we take into account Eqs. \eqref{eq:C1} and \eqref{eq:C2}, we get $h^{\rho\rho}$ and $h^{\phi\phi}$. We can also consider Eq. \eqref{eq:hrrd1} to get $h^{\rho\rho}$, using the first Dirac gauge equation \eqref{eq:dirac1}. Finally, the determinant condition can be used to get $h^{\phi\phi}$, or solve for $C_1$ or $C_2$.\\

A third option, for which all conditions are used, consists in the following steps: solve $h^{zz}$ with Eq. \eqref{eq:hzz}; solve $h^{\rho z}$ with Eq. \eqref{eq:hrzd2}; solve $h^{\rho\rho}$ with Eq. \eqref{eq:hrrd1}; and, finally, get $h^{\phi\phi}$ with Eq. \eqref{eq:det}.\\

These options are beyond the scope of this work, but further investigation is on our future plans. The level of fulfillment of the Dirac gauge and determinant condition may also depend on the specific scenario for which the initial data is computed.\\

\section{\label{sec:con}Conclusions}

In this work we have presented a reformulation of the FCF, including modifications both in the elliptic and the hyperbolic sectors, by introducing two new variables, $V^i$ and $\dot X^i$. With this new reformulation, we keep the local uniqueness properties of the elliptic sector, and, moreover, the new set of equations are presented with a hierarchical structure in terms of PNE. The addition of the new variables also simplifies the source terms of the evolution equations.

In order to numerically test this reformulation, we have  computed stationary initial data of a rotating neutron star. We have discretized the spatial derivatives by means of second-order finite-differences. We have compared our results to the numerical solution using the xCFC scheme, and also to the reference solution obtained with the spectral code LORENE. We have also checked the convergence of our numerical solutions, getting the expected second-order. The solution of the reformulation of the FCF deviates from the one using xCFC as expected: the differences of the variables in the two approaches are of the order of the $h^{ij}$ tensor (or smaller), as it was theoretically established in \cite{cordero2009improved}. Due to the simplification in some equations, some metric variables are solved with more accuracy; in particular, considering the new vector $V^i$ allows one to compute the $\beta^i$ vector more accurately.

The generalized Dirac gauge and the condition on the determinant for the conformal metric have also been checked. To our knowledge, this is the first time when the condition for the determinant has been checked in any finite-differences code for the resolution of Einstein equations in complex non-analytical spacetimes, either in constrained formulations or in free evolution schemes. Although the generalized Dirac gauge is reasonably satisfied (see Figure \ref{im:dirac} for more details), we plan to check in the future the proposed strategies to explicitly impose these restrictions.\\

This new reformulation can indeed be used to compute initial data using finite-differences beyond the xCFC condition, i.e., satisfying all the constraint equations without imposing the conformal metric to be flat.\\

Future steps include complex, highly dynamical simulations solving the hyperbolic sector of the new reformulation and extracting the corresponding gravitational radiation from the system. Another potential application of this work is to use a simplified version of our proposed equations in the context of cosmological simulations by considering only the leading terms of the PNE.

\begin{acknowledgments}
The authors are greatful to J\'er\^ome Novak for fruitful discussions. The authors acknowledge support by the Spanish Agencia Estatal de Investigaci\'on / 
Ministerio de Ciencia, Innovaci\'on y Universidades through the Grants No. 
PGC2018-095984-B-I00, PID2021-125485NB-C21 and the Red Espa\~nola de F\'isica de Ondas Gravitacionales RED2024-153735-E, and by the Generalitat Valenciana through the Grants No. 
PROMETEO/2019/071, CIPROM/2022/49 and ACIF/2019/169 -- European Social Fund. This work has been supported by the European Horizon Europe staff exchange (SE) programme HORIZON-MSCA-2021-SE-01 Grant No. NewFunFiCO-101086251. This research was partially supported by the Perimeter Institute for Theoretical Physics through the Simons Emmy Noether program. Research at Perimeter Institute is supported by the Government of Canada through the Department of Innovation, Science and Economic Development and by the Province of Ontario through the Ministry of Research and Innovation.

\end{acknowledgments}

\sloppy
\bibliographystyle{apsrev4-2}
\bibliography{article}

\end{document}